\documentclass[a4paper,11pt]{article}
\usepackage{pos}

\newcommand{\td}[3]{\frac{d^{#3} #1}{d {#2}^{#3}}} %total derivative of #1 w.r.t #2 of order #3
\renewcommand{\bar}[1]{\ensuremath{\overline{#1}}}

\title{Multi-frequency probes of 2HDM+S dark matter}
\author*{Geoff Beck}

\affiliation{School of Physics and Centre for Astrophysics, University of the Witwatersrand,
	Johannesburg, Wits 2050, South Africa.}

\emailAdd{geoffrey.beck@wits.ac.za}

\abstract{The two-Higgs-doublet with additional scalar (2HDM$+S$) model is one proposed to account for several anomalies that have persisted and increased in significance over runs 1 and 2 at the Large Hadron Collider (LHC). In addition to this, 2HDM+$S$ also supplies a potential Dark Matter (DM) candidate coupling to the Standard Model via the $S$ boson. So far, this model has been difficult to constrain by indirect means. Here we will explore the potential of Omega Centauri, a nearby globular cluster to constrain this interesting DM model. Although such structures are generally considered to be lacking in DM, arguments have been made that this cluster is in fact the relic of a tidally stripped dwarf galaxy. In such a scenario, the DM content would be significant. Combined with its nearness, this would suggest a potential for powerful indirect dark matter signals. We employ both Fermi-LAT gamma-ray data, as well as MeerKAT telescope sensitivities to determine the current status of Omega Centauri as a source of indirect constraints on Weakly Interacting Massive Particles (WIMPs) in a 2HDM+$S$ scenario and for general annihilation channels.}

\FullConference{%
  
High Energy Astrophysics in Southern Africa 2022 - HEASA2022\\
  28 September - 1 October 2022\\
  Brandfort, South Africa
}

\begin{document}
\maketitle

\section{Introduction}
Indirect searches for DM have long been mainly directed towards our own galactic centre (see \cite{Ackermann_2017,Di_Mauro_2021b} and references therein) and an array of Milky-Way satellite galaxies (see e.g.~\cite{Albert_2017,Regis_2021}). So far, these studies have managed to rule out thermal WIMPs, annihilating via $b$-quarks, with masses up to a few hundred GeV. This remains far from closing the window on viable WIMP models. Thus, it makes sense to consider other potential targets. One such possibility would be globular clusters that are the cores of disrupted/stripped dwarf galaxies. An example of such an object may be Omega Centauri, where considerable interest has gone into determining its origin, much of which suggests that the core-remnant scenario is correct~\cite{oc-dsph-2,oc-dsph-3,oc-dsph-4} (not all studies are so positive, see e.g.~\cite{Johnson_2020}). In this scenario Omega Centauri would have a considerable DM component that would survive the tidal stripping, in contrast to regular globular clusters which demonstrate a relative dearth of DM~\cite{Ibata_2013,1996IAUS..174..303H}. This, combined with comparatives nearness to Earth, would result in a very strong annihilation signal from DM. Indeed, it has recently been suggested that DM annihilation could explain observed gamma-ray emissions within this object~\cite{Brown_2019}. With modelling done in \cite{Evans_2022} and \cite{Carlberg_2021} offering evidence in favour of highly dense DM component in Omega Centauri. However, the authors in \cite{Reynoso_Cordova_2022} do not find significant evidence for a DM component in this object.

The 2HDM+$S$ DM model extends the standard model via the addition of a heavy Higgs doublet and a massive neutral scalar. More details of the model and the anomalies can be found in \cite{Fischer:2021sqw} (and references therein). The consequences of a DM candidate coupling to the $S$ boson have been explored in dwarf galaxies at radio frequencies in \cite{beck-atlas2021}. 

In this work we examine the potential of Omega Centauri to produce constraints on WIMP DM by making use of the fitting from \cite{Reynoso_Cordova_2022} to the DM halo parameters. Our findings suggest that Omega Centauri has great potential as a probe of WIMPs. Being able (in an optimistic scenario) to rule out annihilation via $b$-quarks at the thermal relic cross-section for WIMP masses up to 1 TeV. This is through a combination of extant Fermi-LAT data and forecasting with the MeerKAT sensitivity. In addition, Omega-Centauri can potentially rule out the 2HDM+$S$ DM scenario for the  parameter space which explains several astrophysical anomalies (see \cite{beck-atlas2021} for details). These results suggest that further study of Omega Centauri's dynamics are vital, as it potentially presents the most powerful opportunity for indirect DM searches to date. 

This paper is structured as follows: in section~\ref{sec:emm} we layout the formalism for predicting electromagnetic signals from WIMP annihilations, with our results being presented and discussed in section~\ref{sec:res}. Finally, our conclusions are drawn in section~\ref{sec:conc}. 

\section{Gamma-ray \& radio emission from DM annihilation}
\label{sec:emm}
Here we describe the formalism for predicting emissions from WIMP annihilations. 
\subsection{Gamma-rays}
\label{sec:gamma}
To determine the flux of gamma-rays from DM annihilation within a cone subtending $\Delta\Omega$ around the line of sight we use:
\begin{equation}
	S_{\gamma} (E,\Delta\Omega) = \psi(E) \times J(\Delta\Omega) \; , \label{eq:gamma}
\end{equation} 
where the two factors are defined by
\begin{align}
	J(\Delta\Omega) & = \int_{\Delta\Omega} d \Omega \int d z \, \rho_\chi^2 (r) \; , \\
	\psi(E) & = \frac{1}{2} \frac{\langle \sigma V \rangle}{M_\chi^2} \td{N_\gamma}{E}{} (M_\chi,E) \; , \label{eq:j}
\end{align}
where $d\Omega = 2\pi \sin\theta d\theta$, $z$ is the line of sight coordinate, $\rho_\chi$ is the DM density, $r = \sqrt{z^2 +d_L^2 - 2 z d_L \cos\theta}$ is the spherical radius, $d_L$ is the halo luminosity distance, $\langle \sigma V \rangle$ is the thermally averaged DM annihilation cross-section, $M_\chi$ is the DM mass, and $\td{N_\gamma}{E}{} (M_\chi,E)$ are the DM annihilation photon yield functions from \cite{beck-atlas2021} (2HDM+$S$) and \cite{Cirelli_2011,Ciafaloni_2011} ($b$-quark channel). The limits of $z$ integration are $z_\pm = d_L\cos\theta \pm \sqrt{r_t^2 - d_L^2 \sin^2\theta}$, where $r_t = 47$ arcminutes is the tidal radius~\cite{tidalr1975,tidalr1988}. In this work we make use of an Navarro-Frenk-White density profile~\cite{nfw1996}
\begin{equation}
	\rho_{\mathrm{NFW}} (r) = \frac{\rho_s}{\frac{r}{r_s}\left(1+\frac{r}{r_s}\right)^2} \; ,
\end{equation}
where $\rho_s$ and $r_s$ are the characteristic density and radius respectively.

\subsection{The J-factor in Omega Centauri}
\begin{figure}[ht!]
	\centering
	\resizebox{0.49\hsize}{!}{\includegraphics{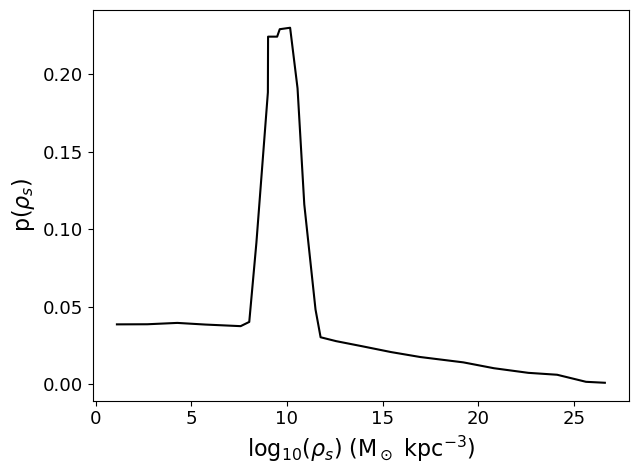}}
	\resizebox{0.49\hsize}{!}{\includegraphics{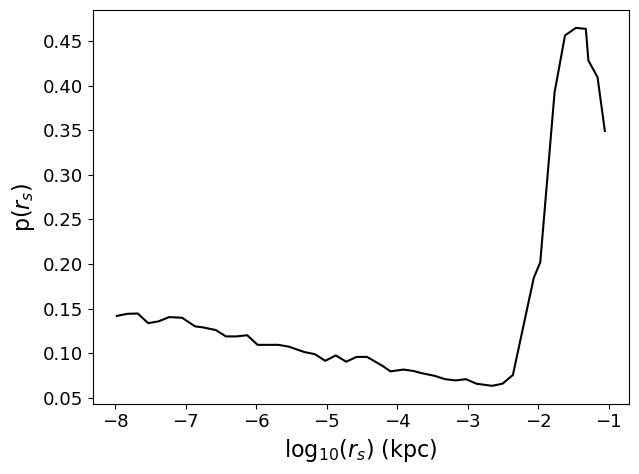}}
	%\resizebox{0.32\hsize}{!}{\includegraphics{j_factor.png}}
	\caption{Distributions for DM parameters in Omega Centauri assuming an NFW profile from \cite{Reynoso_Cordova_2022}. \textit{Left}: $\rho_s$. \textit{Middle}: $r_s$.}
	\label{fig:omega-cen-dm}
\end{figure}

In the recent work \cite{Evans_2022}, the authors fit $r_s$ and $\rho_s$ to optical data sets from the Gaia EDR3 catalogue as well as those from~\cite{Baumgardt_2018,Vasiliev_2021}, and the Hubble space telescope. The resultant range for the J-factor is $10^{22}$ to $10^{24}$ GeV$^2$ cm$^{-5}$ for $\Delta \Omega \geq 0.1$ degrees. This is very similar to the range from the earlier work \cite{Brown_2019}. This suggests the Omega Centauri has a J-factor that is 2-3 orders of magnitude in excess of the most promising dwarf galaxies. However, in \cite{Reynoso_Cordova_2022} the authors discuss the maximum DM content of a variety of globular cluster targets. This is done via the use of line-of-sight stellar velocities sourced from \cite{Kamann_2017,Baumgardt_2018}. They find no significant evidence for a dominant DM component. The results from \cite{Reynoso_Cordova_2022} for the DM parameters in Omega Centauri are reproduced here in Fig.~\ref{fig:omega-cen-dm} (for an NFW profile, but note that Burkert is very similar). The best-fitting values of $r_s$ and $\rho_s$ correspond, via Eq.~(\ref{eq:j}), to $J \sim 10^{21}$ GeV$^2$ cm$^{-5}$ for $\Delta \Omega \geq 0.1$ degrees. The difference to \cite{Evans_2022} is argued by the authors of \cite{Reynoso_Cordova_2022} to be down to the former work using a more restricted set of parameters to describe the stellar component. 

Taking this into account we will consider the range $10^{20}$ to $10^{24}$ GeV$^2$ cm$^{-5}$ but note that this may be optimistic. This is because \cite{Reynoso_Cordova_2022} did not find evidence of a dominant DM component. However, it must be noted that others works~\cite{Carlberg_2021} have indicated that evidence for a significant DM component emerges at radii larger than those considered by \cite{Reynoso_Cordova_2022}. This means that, while the conflicting analysis of \cite{Reynoso_Cordova_2022} should caution us against optimistic J-factor values, there is substantial statistical room for Omega Centauri to be strongly DM dominated.

\subsection{Radio emission}
To compute radio emission we determine the surface brightness a distance $R$ from the centre of the halo as
\begin{equation}
	I_{\mathrm{sync}} (\nu,R) = \int dl \, \frac{j_{\mathrm{sync}}(\nu,\sqrt{R^2+l^2})}{4 \pi} \; , 
	\label{eq:sb}
\end{equation}
where $l$ is the line-of-sight coordinate and the integral runs over the line of sight through the target at $R$. The emissivity $j$ is given by
\begin{equation}
	j_{\mathrm{sync}} (\nu,r) = \int_{m_e}^{M_\chi} dE \, \td{n_{e^\pm}}{E}{}(E,r) P_{\mathrm{sync}} (\nu,E,r) \; ,
	\label{eq:emm}
\end{equation}
where $P_{\mathrm{sync}}$ is the synchrotron power and $\td{n_{e^\pm}}{E}{}(E,r)$ is the solution of diffusion-loss equation with DM annihilation as a source (for details see \cite{Beck_2023}). For radio computations we choose two sets of $r_s$ and $\rho_s$ values. First we use $30$ pc and $2.5\times 10^9$ M$_\odot$ kpc$^{-3}$ to match $J \sim 10^{20}$ GeV$^2$ cm$^{-5}$. Second, we use $30$ pc and $2.5\times 10^{11}$ M$_\odot$ kpc$^{-3}$ to match $J \sim 10^{24}$ GeV$^2$ cm$^{-5}$. These choices are consistent with the peaks in Fig.~\ref{fig:omega-cen-dm}. We make use of the median turbulent Milky-Way magnetic field model at Omega Centauri, derived by \cite{Kar_2020}, with $B(r) = 5$ $\mu$G and diffusion constant $D_0 = 3\times 10^{26}$ cm$^2$ s$^{-1}$.

%\begin{figure}[ht!]
%	\centering
%	\resizebox{0.7\hsize}{!}{\includegraphics{j_factor.png}}
%	\caption{J-factor distribution deduced from DM parameters of \cite{Reynoso_Cordova_2022}.}
%	\label{fig:omega-cen-j}
%\end{figure}

\section{DM constraints}
\label{sec:res}
\begin{figure}[ht!]
	\centering
	\resizebox{0.49\hsize}{!}{\includegraphics{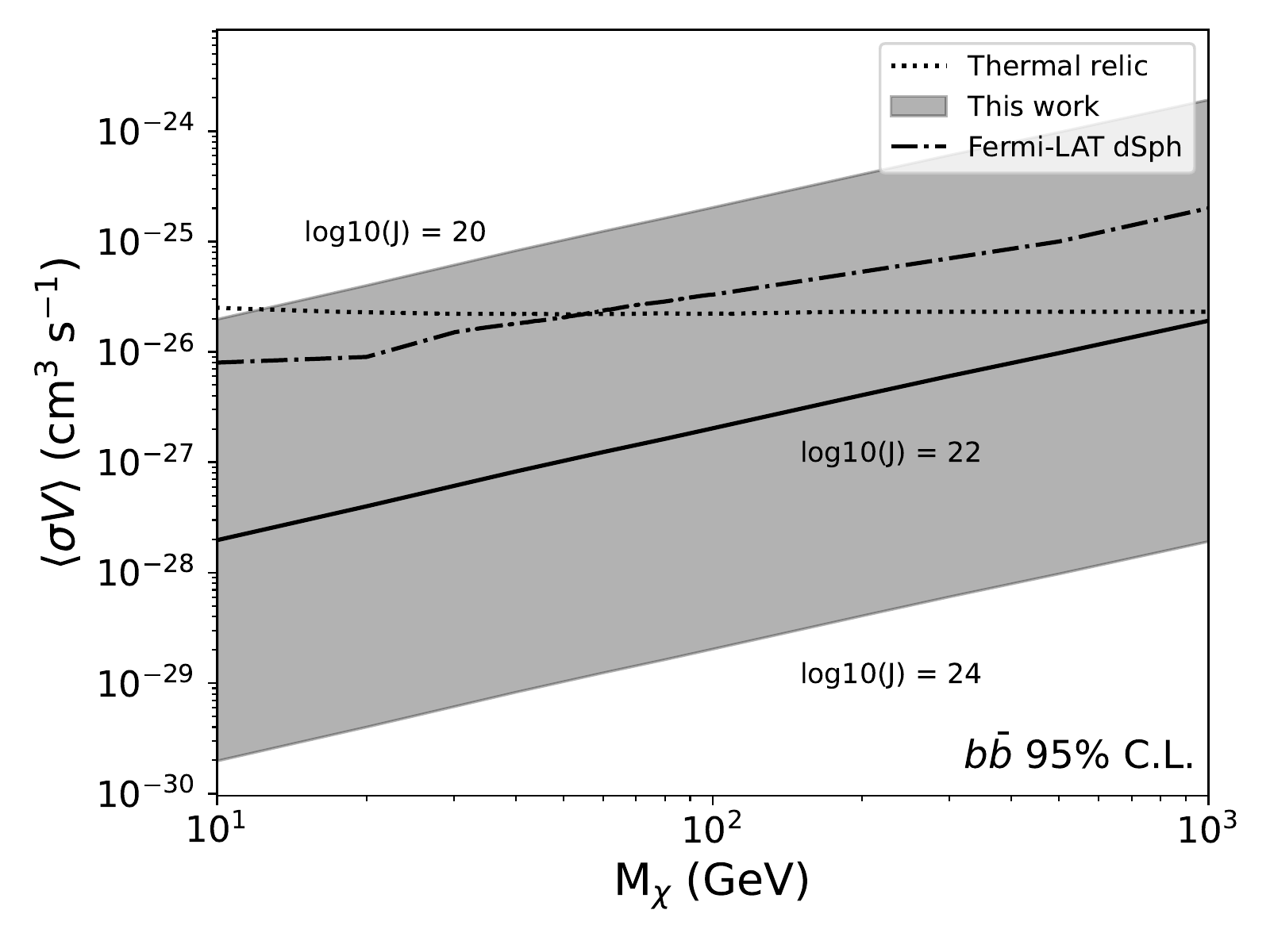}}
	\resizebox{0.49\hsize}{!}{\includegraphics{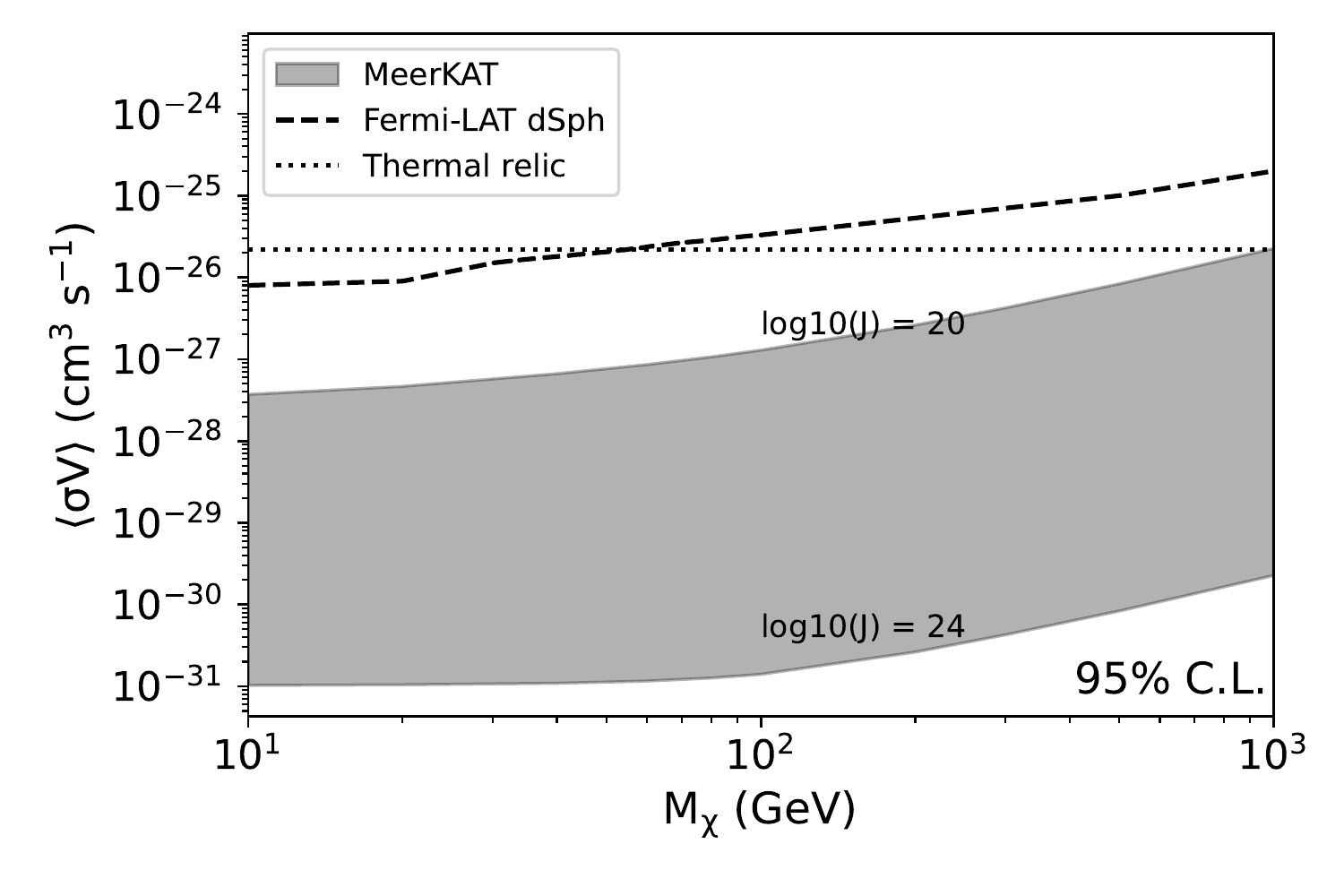}}
	\caption{Constraints on $\langle \sigma V \rangle$, for annihilation via $b\bar{b}$, at 95\% confidence interval. Left: using Fermi-LAT data~\cite{Brown_2019}. Right: using MeerKAT sensitivities~\cite{beck-makh-2022}.}
	\label{fig:bb}
\end{figure}

In Fig.~\ref{fig:bb} we display the 95\% confidence interval limits on $\langle \sigma V\rangle$, via $b\bar{b}$ channel, from both Fermi-LAT data~\cite{Brown_2019} and MeerKAT sensitivities~\cite{beck-makh-2022}. In the case of Fermi-LAT data, we compare to limits from dwarf spheroidal galaxies~\cite{Albert_2017}. These turn out to be slightly superior to the minimal J-factor scenario in Omega Centauri. However, it is clear that Omega Centauri has strong potential as a target for indirect DM searches, with J-factors $\geq 10^{22}$ GeV$^2$ cm$^{-5}$ able to exclude the thermal relic cross-section out to at least 1 TeV. Despite this, the J-factor uncertainty suggests that these results must be treated with caution. 

In the case of MeerKAT projections (for 20 hours of observation time) we see the potential to rule out thermal WIMPs at masses up to 1 TeV for J-factors $\geq 10^{20}$ GeV$^2$ cm$^{-5}$. This substantially stronger than for Fermi-LAT. However, it does not account for magnetic field uncertainties. These are likely to be smaller than the J-factor uncertainties~\cite{Kar_2020}. Notably, the most optimistic case for Omega Centauri may rule out the WIMP parameter space entirely.

\begin{figure}[ht!]
	\centering
	\resizebox{0.49\hsize}{!}{\includegraphics{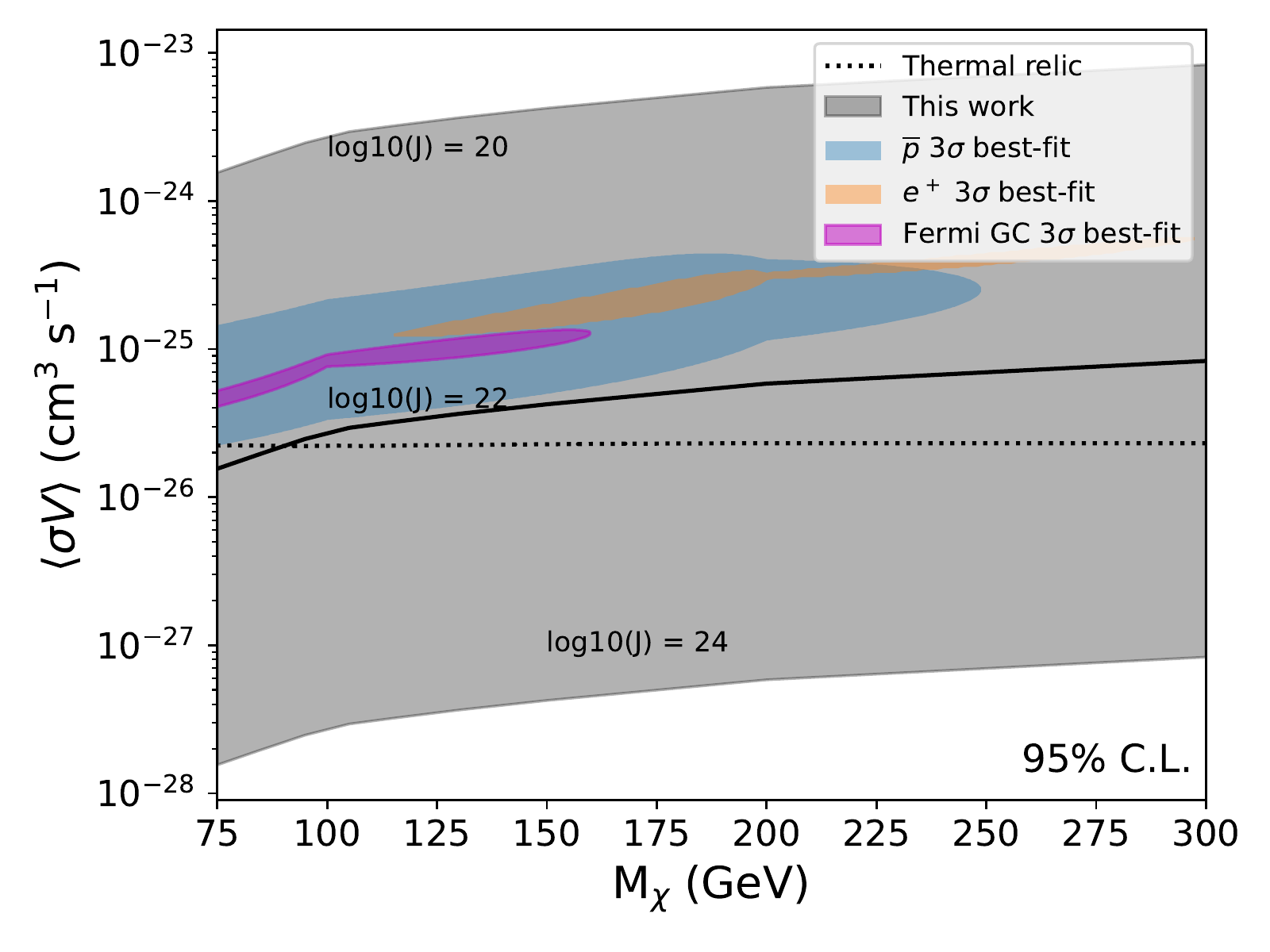}}
	\resizebox{0.49\hsize}{!}{\includegraphics{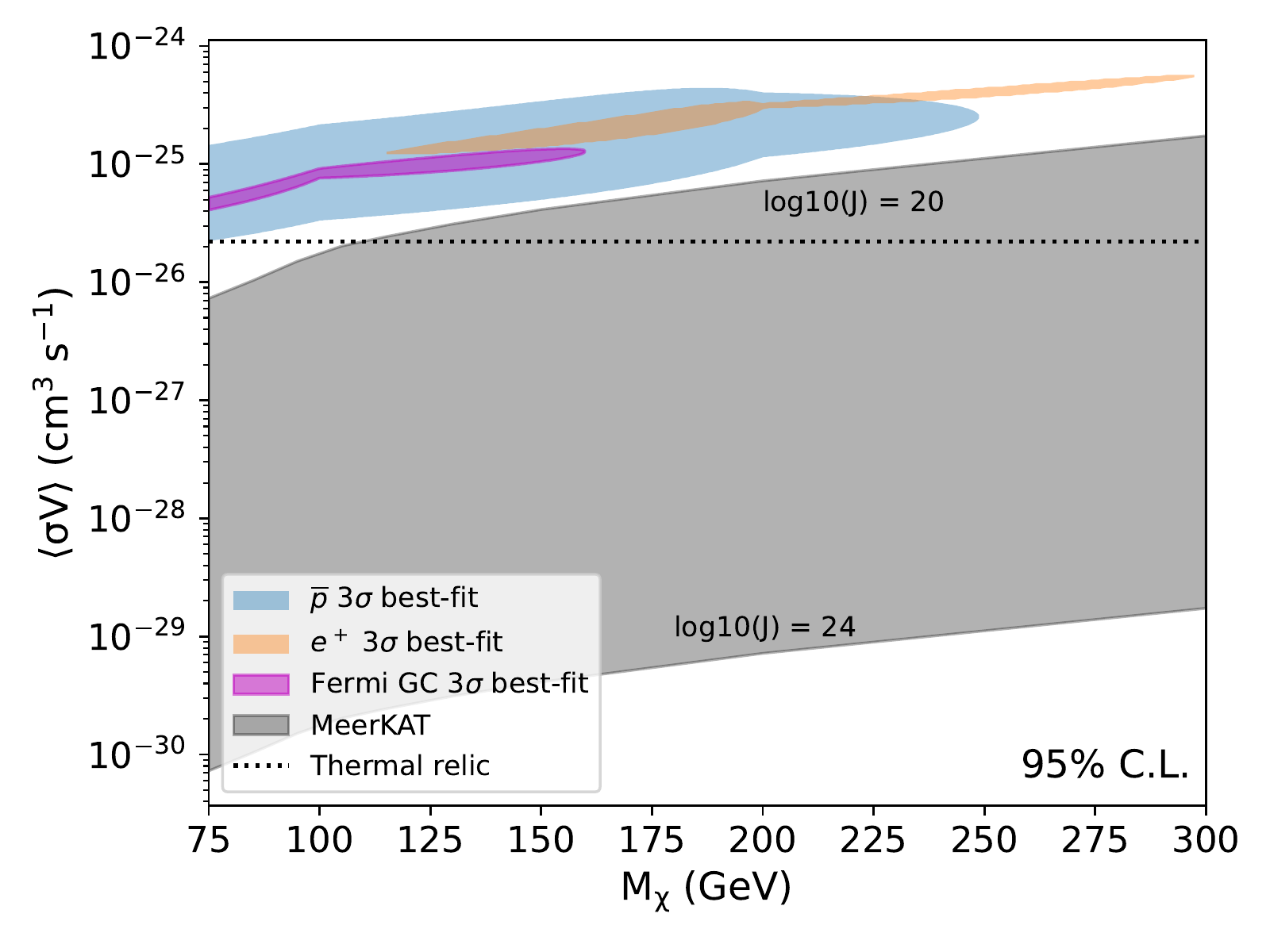}}
	\caption{Constraints on $\langle \sigma V \rangle$ for 2HDM+$S$ DM at 95\% confidence interval. See \cite{beck-atlas2021} for details of the various shaded regions. Left: using Fermi-LAT data~\cite{Brown_2019}. Right: using MeerKAT sensitivities~\cite{beck-makh-2022}.}
	\label{fig:2hdms}
\end{figure}

In Fig.~\ref{fig:2hdms} we display the 95\% confidence interval limits on $\langle \sigma V\rangle$, via the 2HDM+$S$ channel, from Fermi-LAT data~\cite{Brown_2019} and MeerKAT sensitivities~\cite{beck-makh-2022}. The required J-factor to achieve complete exclusion of the parameter space corresponding to astrophysical excess~\cite{beck-atlas2021} is $\sim 10^{22}$ GeV$^2$ cm$^{-5}$. However, in more optimistic scenarios it is both possible to probe below the relic limit and to fully exclude the best-fitting excess parameter space. This is significant because Fermi-LAT dwarf galaxy data is insufficient to constrain this parameter space~\cite{beck-atlas2021}, but performs slightly better than the minimum considered J-factor for Omega Centauri. 

For MeerKAT projections we can exclude the entire relevant parameter space with J-factors $\geq 10^{20}$ GeV$^2$ cm$^{-5}$ and can probe well below the relic level in the most optimistic scenarios for masses out to several hundred GeV at least.

\section{Conclusions}
\label{sec:conc}
In optimistic scenarios, Omega Centauri's large J-factor allows us to probe below the thermal relic WIMP cross-section for masses up to 1 TeV, through a combination of existing Fermi-LAT data and forecasting for MeerKAT. This applies for both generic WIMPs and those from 2HDM+$S$. It must be noted, however, that there is considerable uncertainty in the estimation of the J-factor in Omega Centauri. Some authors find no evidence for a significant DM component~\cite{Reynoso_Cordova_2022}, while others suggest the opposite~\cite{Brown_2019,Evans_2022,Carlberg_2021}. In light of this vast potential to rule out thermal WIMP DM, further studies are required to more precisely determine the DM content of Omega Centauri.   

\bibliographystyle{iopart-num}
\bibliography{heasa2022}

\end{document}